\documentclass[10pt]{article}
\usepackage{graphicx}
\usepackage{amsmath}
\usepackage{amssymb}
\usepackage{caption2}
\begin{document}
\bibliographystyle{prsty}
\begin{center}
{\large {\bf \sc{  Analysis of the $Z_{cs}(3985)$ as  the axialvector tetraquark  state  }}} \\[2mm]
Zhi-Gang Wang \footnote{E-mail: zgwang@aliyun.com.  }    \\
 Department of Physics, North China Electric Power University, Baoding 071003, P. R. China
\end{center}

\begin{abstract}
In this paper, we choose the scalar and axialvector diquark operators in the color antitriplet as the fundamental  building blocks to construct the four-quark  currents and   investigate   the diquark-antidiquark type axialvector tetraquark states $c\bar{c}u\bar{s}$  in the framework of  the QCD sum rules. The predicted tetraquark  mass $M_Z=3.99\pm0.09\,\rm{GeV}$ is in excellent agreement with the experimental value  $3985.2^{+2.1}_{-2.0}\pm1.7\,\rm{MeV}$ from the BESIII collaboration, which  supports identifying  the $Z_{cs}(3985)$   as the cousin of the $Z_c(3900)$ with the quantum numbers  $J^{PC}=1^{+-}$. We take into account the light flavor   $SU(3)$ mass-breaking effect to estimate the mass spectrum of the diquark-antidiquark type hidden-charm  tetraquark states having  the strangeness according to  the previous works.
\end{abstract}



\section{Introduction}

Very recently, the BESIII collaboration observed an excess over the known contributions of the conventional charmed mesons  near the $D_s^- D^{*0}$ and $D^{*-}_s D^0$ mass thresholds in the $K^{+}$ recoil-mass spectrum with the significance of  5.3 $\sigma$ in the processes of the $e^+e^-\to K^+ (D_s^- D^{*0} + D^{*-}_s D^0)$ \cite{BES3985}.  It is the first candidate of the charged hidden-charm tetraquark state with  strangeness.
 The Breit-Wigner  mass and width of the new structure, $Z_{cs}(3985)$, were determined to be  $3985.2^{+2.1}_{-2.0}\pm1.7\,\rm{MeV}$   and $13.8^{+8.1}_{-5.2}\pm4.9\,\rm{MeV}$, respectively. According to the production mode, it is nature to draw the conclusion that the $Z_{cs}(3985)$ should be a cousin of the well-known $Z_c(3885/3900)$  with  strangeness, the $Z_c(3885)$ was observed in the precess  $e^{+}e^{-} \to (D\bar{D}^{*})^-\pi^+$,  which means that the $Z_c(3885)$ and
$Z_{cs}(3985)$ are governed by a similar production mechanism and have a similar quark structure $c\bar{c}q\bar{u}$ with the $q=d$ or $s$.
The discovery of the $Z_{cs}(3985)$ could provide some unique hints to uncover the secrets of the charged exotic $Z_c$ structures.

The existence of an exotic  $Z_{cs}$ state with a mass,  which lies near the $D_s^- D^{*0}$ and $D^{*-}_s D^0$ thresholds $3975.2\,\rm{MeV}$ and $3977.0\,\rm{MeV}$ respectively,  has been predicted in several theoretical models, including the diquark-antidiquark type  tetraquark model \cite{Ebert:2008kb, Ferretti:2020ewe}, the $D_s\bar{D}^*$ molecule  model~\cite{Lee:2008uy,Dias:2013qga}, the hadro-quarkonium model~\cite{Ferretti:2020ewe,SB-Voloshin}, and  the initial-single-chiral-particle-emission mechanism~\cite{Chen:2013wca}.
Just like the $Z_c$ states, the decay rate of the $Z_{cs}$ to open-charm final states is expected to be larger than the decay rate to charmonium final states.

After the discovery of the $Z_{cs}(3985)$, several possible explanations for its nature were proposed,  such as
the tetraquark state \cite{Zcs-CFQiao,Zcs-tetra-quark},  $D^* \bar{D}_s-D \bar{D}_s^*$ hadronic molecular state or dynamically  generated resonance with the  coupled-channel effects \cite{Zcs-CFQiao,Zcs-molecule-1,Zcs-molecule-2,Zcs-molecule-3,Zcs-molecule-4,Zcs-molecule-5,Zcs-molecule-6}, re-scattering effects  \cite{Zcs-Rescatt-1,Zcs-Rescatt-2}, and we can study its nature with the photo-production \cite{Zcs-Photopro}.

In Ref.\cite{WangHuangTao-3900},  we  investigate the  spin-parity $J^P=1^+$ hidden-charm  tetraquark states without strangeness in the framework of  the QCD sum rules,
and examine the dependence of the masses and pole residues on the energy scales $\mu$ of the spectral densities at the quark-gluon level at length  for the first time. Our reliable calculations support identifying the $X(3872)$ and $Z_c(3900)$  as  the  diquark-antidiquark type tetraquark states having hidden-charm  with the quantum numbers  $J^{PC}=1^{++}$ and $1^{+-}$, respectively.
In Ref.\cite{Wang-Hidden-charm}, we take the diquark and antidiquark operators in the color antitriplet and color triplet respectively  as the fundamental  building blocks to construct
  the scalar, axialvector and tensor  four-quark local currents to investigate the  mass spectrum of the ground state  tetraquark states with hidden-charm but without strangeness  in the framework of
the QCD sum rules  comprehensively, and  revisit the possible identifications of the existing tetraquark candidates with hidden-charm, such as the  $X(3860)$, $X(3872)$, $X(3915)$,  $X(3940)$, $X(4160)$, $Z_c(3900)$, $Z_c(4020)$, $Z_c(4050)$, $Z_c(4055)$, $Z_c(4100)$, $Z_c(4200)$, $Z_c(4250)$, $Z_c(4430)$, $Z_c(4600)$, etc.

In Refs.\cite{WangHuangTao-3900,Wang-Hidden-charm}, we choose the local  axialvector four-quark currents $\eta^N_\mu(x)$ and $\eta^P_\mu(x)$,
\begin{eqnarray}
\eta^N_{\mu}(x)&=&\frac{\varepsilon^{ijk}\varepsilon^{imn}}{\sqrt{2}}\Big\{u^{T}_j(x)C\gamma_5 c_k(x) \bar{d}_m(x)\gamma_\mu C \bar{c}^{T}_n(x)-u^{T}_j(x)C\gamma_\mu c_k(x)\bar{d}_m(x)\gamma_5 C \bar{c}^{T}_n(x) \Big\} \, , \nonumber\\
\eta^P_{\mu}(x)&=&\frac{\varepsilon^{ijk}\varepsilon^{imn}}{\sqrt{2}}\Big\{u^{T}_j(x)C\gamma_5 c_k(x) \bar{d}_m(x)\gamma_\mu C \bar{c}^{T}_n(x)+u^{T}_j(x)C\gamma_\mu c_k(x)\bar{d}_m(x)\gamma_5 C \bar{c}^{T}_n(x) \Big\} \, ,
\end{eqnarray}
to investigate    the lowest tetraquark states with the quantum numbers $J^{PC}=1^{+-}$ and $1^{++}$, respectively, where the $i$, $j$, $k$, $m$, $n$ are color indices.  The four-quark current $\eta^N_\mu(x)$ has the quantum numbers $J^{PC}=1^{+-}$ and couples potentially to the $Z_c^+(3900)$,
in fact, the $Z_c^\pm(3900)$ have non-zero electric charge, and do not have definite charge conjugation, or they are not eigenstates  of the charge conjugation,   only the electric neutral state $Z_c^0(3900)$ has definite charge conjugation.
If the $Z_{cs}^-(3985)$ is really  a tetraquark state, irrespective of whether   it is the diquark-antidiquark type or meson-meson type tetraquark state, its valence quarks  are
$c\bar{c}s\bar{u}$, and  has no definite electric conjugation indeed (to be more precisely), we suppose that it has definite conjugation, just like its cousins $c\bar{c}q\bar{q}$ and $c\bar{c}s\bar{s}$.
In the present work, we tentatively identify  the $Z_{cs}^-(3985)$ as the diquark-antidiquark type axialvector tetraquark state with the four   valence quarks $c\bar{c}s\bar{u}$ and examine  its mass in the framework of  the QCD sum rules at length. Then we take into account the light flavor $SU(3)$ mass-breaking effect to  explore the
mass spectrum of the diquark-antidiquark type tetraquark states with hidden-charm and with strangeness according to our previous works.

 The article is arranged as follows:  in Sect.2, we get the QCD sum rules for the masses and pole residues of  the
$J^P=1^+$ tetraquark  states with hidden-charm and with strangeness;  in Sect.3, we obtain  numerical results and give discussions; and Sect.4 is aimed to get a
conclusion.

\section{The  QCD sum rules for the axialvector tetraquark states with strangeness}
If we choose the favorable diquark  configurations,  the scalar ($S$) and axialvector ($A$) diquark states in the color antitriplet,  as the fundamental building blocks to construct the diquark-antidiquark type tetraquark states, we can obtain two nonets with the symbolic  structures,
\begin{eqnarray}\label{1-nonet}
 I=1 &:& [uc]_S[\bar{d}\bar{c}]_A-[uc]_A[\bar{d}\bar{c}]_S\, , \,[dc]_S[\bar{u}\bar{c}]_A-[dc]_A[\bar{u}\bar{c}]_S\, ,\, \nonumber\\ &&\frac{[uc]_S[\bar{u}\bar{c}]_A-[uc]_A[\bar{u}\bar{c}]_S-[dc]_S[\bar{d}\bar{c}]_A+[dc]_A[\bar{d}\bar{c}]_S}{\sqrt{2}}\, ; \nonumber \\
I=0 &:& \frac{[uc]_S[\bar{u}\bar{c}]_A-[uc]_A[\bar{u}\bar{c}]_S+[dc]_S[\bar{d}\bar{c}]_A-[dc]_A[\bar{d}\bar{c}]_S}{\sqrt{2}}\, ,\nonumber\\
     & &[sc]_S[\bar{s}\bar{c}]_A-[sc]_A[\bar{s}\bar{c}]_S\, ; \nonumber\\
I=\frac{1}{2} & :&[qc]_S[\bar{s}\bar{c}]_A-[qc]_A[\bar{s}\bar{c}]_S\, , \,[sc]_S[\bar{q}\bar{c}]_A-[sc]_A[\bar{q}\bar{c}]_S\, ,
\end{eqnarray}
and
\begin{eqnarray}\label{2-nonet}
 I=1 &:& [uc]_S[\bar{d}\bar{c}]_A+[uc]_A[\bar{d}\bar{c}]_S\, , \,[dc]_S[\bar{u}\bar{c}]_A+[dc]_A[\bar{u}\bar{c}]_S\, ,\, \nonumber\\ &&\frac{[uc]_S[\bar{u}\bar{c}]_A+[uc]_A[\bar{u}\bar{c}]_S-[dc]_S[\bar{d}\bar{c}]_A-[dc]_A[\bar{d}\bar{c}]_S}{\sqrt{2}} \, ;\nonumber \\
I=0 &:& \frac{[uc]_S[\bar{u}\bar{c}]_A+[uc]_A[\bar{u}\bar{c}]_S+[dc]_S[\bar{d}\bar{c}]_A+[dc]_A[\bar{d}\bar{c}]_S}{\sqrt{2}}\, ,\nonumber\\
     & &[sc]_S[\bar{s}\bar{c}]_A+[sc]_A[\bar{s}\bar{c}]_S\, ; \nonumber\\
I=\frac{1}{2} & :&[qc]_S[\bar{s}\bar{c}]_A+[qc]_A[\bar{s}\bar{c}]_S\, , \,[sc]_S[\bar{q}\bar{c}]_A+[sc]_A[\bar{q}\bar{c}]_S\, ,
\end{eqnarray}
where the $I=1$, $0$, $\frac{1}{2}$ are the isospins of the hidden-charm tetraquark states, $q=u$, $d$.

In the first nonet, only the charge-neutral tetraquark states $\frac{1}{\sqrt{2}}\Big([uc]_S[\bar{u}\bar{c}]_A-[uc]_A[\bar{u}\bar{c}]_S-[dc]_S[\bar{d}\bar{c}]_A+[dc]_A[\bar{d}\bar{c}]_S \Big)$,
$\frac{1}{\sqrt{2}}\Big([uc]_S[\bar{u}\bar{c}]_A-[uc]_A[\bar{u}\bar{c}]_S+[dc]_S[\bar{d}\bar{c}]_A-[dc]_A[\bar{d}\bar{c}]_S\Big)$ and
     $[sc]_S[\bar{s}\bar{c}]_A-[sc]_A[\bar{s}\bar{c}]_S$ are eigenstates of the charge conjugation operators, and have the definite charge conjugation $C=-$.

 In the second nonet, only the charge-neutral tetraquark states
$\frac{1}{\sqrt{2}}\Big([uc]_S[\bar{u}\bar{c}]_A+[uc]_A[\bar{u}\bar{c}]_S-[dc]_S[\bar{d}\bar{c}]_A-[dc]_A[\bar{d}\bar{c}]_S\Big)$,
$ \frac{1}{\sqrt{2}}\Big([uc]_S[\bar{u}\bar{c}]_A+[uc]_A[\bar{u}\bar{c}]_S+[dc]_S[\bar{d}\bar{c}]_A+[dc]_A[\bar{d}\bar{c}]_S \Big)$ and
     $[sc]_S[\bar{s}\bar{c}]_A+[sc]_A[\bar{s}\bar{c}]_S$ are eigenstates of the charge conjugation operators, and have the definite charge conjugation $C=+$.

Thereafter, we will assume that the first nonet and second nonet have the negative and positive charge conjugations respectively so as to distinguish the two nonets, and we should bear in mind that the charge conjugation is not a good quantum number.

Routinely, let us write down  the two-point vacuum Green's functions $\Pi_{\mu\nu}(p)$,
\begin{eqnarray}
\Pi_{\mu\nu}(p)&=&i\int d^4x e^{ip \cdot x} \langle0|T\left\{J_\mu(x)J^{\dagger}_\nu(0)\right\}|0\rangle \, ,
\end{eqnarray}
where the interpolating currents $J_\mu(x)=J^N_\mu(x)$ and $J^P_\mu(x)$,
\begin{eqnarray}
J^N_{\mu}(x)&=&\frac{\varepsilon^{ijk}\varepsilon^{imn}}{\sqrt{2}}\Big\{u^{T}_j(x)C\gamma_5 c_k(x) \bar{s}_m(x)\gamma_\mu C \bar{c}^{T}_n(x)-u^{T}_j(x)C\gamma_\mu c_k(x)\bar{s}_m(x)\gamma_5 C \bar{c}^{T}_n(x) \Big\} \, , \nonumber\\
J^P_{\mu}(x)&=&\frac{\varepsilon^{ijk}\varepsilon^{imn}}{\sqrt{2}}\Big\{u^{T}_j(x)C\gamma_5 c_k(x) \bar{s}_m(x)\gamma_\mu C \bar{c}^{T}_n(x)+u^{T}_j(x)C\gamma_\mu c_k(x)\bar{s}_m(x)\gamma_5 C \bar{c}^{T}_n(x) \Big\} \, ,
\end{eqnarray}
the superscripts $N$ and $P$ stand for the negative and positive charge conjugations, respectively, the $i$, $j$, $k$, $m$, $n$ are color indices. We suppose that the interpolating four-quark currents $J^N_\mu(x)$ and $J_\mu^P(x)$ have the negative and positive charge conjugation respectively in the sense of the limit $J_\mu(x)|_{s \to u}$ or $J_\mu(x)|_{u \to s}$, which should be understood in the same way as in the two tetraquark nonets in Eqs.\eqref{1-nonet}-\eqref{2-nonet}.
In Refs.\cite{WangHuangTao-3900,Wang-Hidden-charm}, we  choose the four-quark  currents $\eta^N_\mu(x)$ and $\eta^P_\mu(x)$
to explore the axialvector tetraquark states $uc\bar{d}\bar{c}$, and observe that the $uc\bar{d}\bar{c}$ states  with the quantum numbers $J^{PC}=1^{+-}$ and $1^{++}$ have almost degenerated masses. The currents $J^N_\mu(x)$ and $J^P_\mu(x)$ are the $SU(3)$ partners of the currents $\eta^N_\mu(x)$ and $\eta^P_\mu(x)$, respectively with the simple relation $d \leftrightarrow s$.
Now we also expect that the four-quark currents $J^N_\mu(x)$ and $J_\mu^P(x)$ couple potentially to the $J^{PC}=1^{+-}$ and $1^{++}$ tetraquark states with almost the same  masses.

At the hadron  side, we insert  a complete set  of intermediate  tetraquark states with hidden-charm, strangeness and other quantum numbers, such as the spin, parity, charge conjugation,  as the four-quark current operators  $J_\mu(x)$,   into the Green's functions $\Pi_{\mu\nu}(p)$ to obtain the hadron spectral  representation
\cite{SVZ79,PRT85}, and  separate  the  lowest  tetraquark states $Z_{N/P}$ with hidden-charm and with strangeness (or the lowest pole terms), and obtain the
 results:
\begin{eqnarray}\label{CF-Hadron}
\Pi_{\mu\nu}(p) & = & \frac{\lambda_Z^2}{M^2_Z-p^2}\left(-g_{\mu\nu}+\frac{p_{\mu}p_{\nu}}{p^2} \right) +\cdots  \, ,\nonumber\\
&=&\Pi(p^2)\left(-g_{\mu\nu}+\frac{p_{\mu}p_{\nu}}{p^2} \right) +\cdots\, ,
 \end{eqnarray}
 where the pole residues $\lambda_Z$ (in other works, the decay constants)  are defined by $\langle 0|J_\mu(0)|Z_{N/P}(p)\rangle=\lambda_Z\varepsilon_\mu$, the $\varepsilon_\mu$ are the polarization vectors of the axialvector tetraquark states $Z_{N/P}$. On the other hand, there are  two-particle scattering state contributions from the two-meson pairs
$K J/\psi$,  $\eta_c K^*$, $D_s \bar{D}^*$, $D^* \bar{D}_s $, $\cdots$, as the quantum field theory does allow  non-vanishing couplings between the four-quark currents and  two-particle scattering states in case that they share the same quantum numbers.  In Ref.\cite{WZG-IJMPA-two-par}, we investigate  the $Z_c(3900)$ as an axialvector tetraquark state with the quantum numbers   $J^{PC}=1^{+-}$  in the framework of the QCD sum rules  at length  by considering  the two-particle scattering state contributions  and the nonlocal effects between the two colored constituents (diquark and antidiquark) in the four-quark current operator,  and obtain the conclusion that   the contribution of the $Z_c(3900)$ as a pole term plays a non-substitutable role, we can saturate the QCD sum rules at the hadron side no matter  with or without the two-meson scattering state contributions. The net effects of the two-particle scattering states of the intermediate  meson pairs   can be taken into account effectively  by adding an energy-dependent finite width to the pole term. In the present case, the energy-dependent Breit-Wigner   width $13.8^{+8.1}_{-5.2}\pm4.9\,\rm{MeV}$ of the $Z_{cs}(3985)$ is really small enough so as to be neglected safely.

In the QCD side, we carry out the operator product expansion up to the vacuum condensates of dimension 10 in a consistent way, just as what we did in our previous works \cite{WangHuangTao-3900,Wang-Hidden-charm}, in the deep Euclidean region $P^2=-p^2\to\infty$ or $\gg \Lambda^2_{QCD}$, which corresponds to the small  spatial distance and
time interval $\vec{x}\sim t\sim \frac{1}{\sqrt{P^2}}$. We take into account or calculate the vacuum condensates   $\langle\bar{q}q\rangle$,  $\langle\bar{s}s\rangle$, $\langle\bar{q}g_{s}\sigma Gq\rangle$, $\langle\bar{s}g_{s}\sigma Gs\rangle$, $\langle\frac{\alpha_{s}GG}{\pi}\rangle$,  $\langle\bar{q}q\rangle\langle\bar{s}s\rangle $,
$\langle\bar{q}q\rangle \langle\frac{\alpha_{s}GG}{\pi}\rangle$, $\langle\bar{s}s\rangle \langle\frac{\alpha_{s}GG}{\pi}\rangle$,
$\langle\bar{q}q\rangle  \langle\bar{s}g_{s}\sigma Gs\rangle$, $\langle\bar{s}s\rangle  \langle\bar{q}g_{s}\sigma Gq\rangle$,
$\langle\bar{q}q\rangle\langle\bar{s}s\rangle \langle\frac{\alpha_{s}GG}{\pi}\rangle$ and $\langle\bar{q}g_{s}\sigma Gq\rangle\langle\bar{s}g_{s}\sigma Gs\rangle$, which are the vacuum expectation values of the quark-gluon operators of the orders $\mathcal{O}(\alpha_s^k)$ with the restriction $k\leq1$ as $\alpha_s=\frac{g_s^2}{4\pi}$.
The vacuum condensates $\langle g_s^3f_{abc}G^aG^bG^c\rangle$, $\langle\frac{\alpha_sGG}{\pi}\rangle^2$, $\langle\frac{\alpha_sGG}{\pi}\rangle \langle \bar{q}g_s\sigma G q\rangle$ and $\langle \bar{q}q\rangle\langle g_s^3f_{abc}G^aG^bG^c\rangle$ are of the dimensions  $6$, $8$, $9$ and $9$, respectively, and are vacuum expectation values of the quark-gluon operators of the orders $\mathcal{O}(\alpha_s^{\frac{3}{2}})$, $\mathcal{O}(\alpha_s^{2})$, $\mathcal{O}(\alpha_s^{\frac{3}{2}})$  and $\mathcal{O}(\alpha_s^{\frac{3}{2}})$, respectively, and are  neglected in the present work,  just like in our previous works \cite{Wang-Hidden-charm,Wang-tetra-formula,WangZG-eff-Mc,Wang-tetra-IJMPA}, direct calculations indicate that those contributions are tiny indeed \cite{WangXW-Wang}.
In calculations, we have assumed  vacuum saturation for the sake of factorizing the higher dimensional vacuum condensates into the lowest ones to reduce the numbers of the fundamental parameters, which works very well indeed in the large color numbers limit.

Once we get the analytical expressions of the correlation (or Green's) functions $\Pi(p^2)$ at the  degrees of freedom of the quarks and gluons, then we resort to dispersion relation to get the spectral densities at the quark level straightforwardly, and   match the two sides of the correlation (Green's) functions $\Pi(p^2)$ (i.e. the hadron side and QCD side),  accomplish  the quark-hadron duality (in other words, the current-hadron duality) below the thresholds $s_0$ of the  continuum states or higher resonances, complete the Borel transform in regard  to the variable or parameter $P^2=-p^2$  and  acquire  the  QCD sum rules:
\begin{eqnarray}\label{QCDSR}
\lambda_{Z}^2\exp\left( -\frac{M_{Z}^2}{T^2}\right)&=& \int_{4m_c^2}^{s_0}ds \,\rho_{QCD}(s)\,\exp\left( -\frac{s}{T^2}\right)\,  ,
\end{eqnarray}
where  the explicit expressions of the   spectral densities $\rho_{QCD}(s)$ at the quark level are neglected for simplicity.

We differentiate   both sides of the above equation  in regard  to the parameter  $\frac{1}{T^2}$, then eliminate the
 pole residues $\lambda_{Z}$ by introducing  a  fraction,  and reach  the QCD sum rules for
 the masses of the hidden-charm axialvector tetraquark states with  strangeness,
 \begin{eqnarray}
 M^2_{Z} &=& \frac{-\int_{4m_c^2}^{s_0}ds \frac{d}{d(1/T^2)}\, \rho_{QCD}(s)\,\exp\left( -\frac{s}{T^2}\right)}{\int_{4m_c^2}^{s_0}ds \, \rho_{QCD}(s)\,\exp\left( -\frac{s}{T^2}\right)}\,  .
\end{eqnarray}

\section{Numerical results and discussions}
We adopt  the standard values or conventional values of all the  vacuum condensates
$\langle\bar{q}q \rangle=-(0.24\pm 0.01\, \rm{GeV})^3$,  $\langle\bar{q}g_s\sigma G q \rangle=m_0^2\langle \bar{q}q \rangle$,  $\langle\bar{s}s \rangle=(0.8\pm0.1)\langle\bar{q}q \rangle$,
$\langle\bar{s}g_s\sigma G s \rangle=m_0^2\langle \bar{s}s \rangle$,
$m_0^2=(0.8 \pm 0.1)\,\rm{GeV}^2$, $\langle \frac{\alpha_s
GG}{\pi}\rangle=0.012\pm0.004\,\rm{GeV}^4$    at the typical energy scale  $\mu=1\, \rm{GeV}$
\cite{SVZ79,PRT85,Ioffe-NPB-1981,Ioffe-mixcondensate,ColangeloReview}, and  prefer the $\overline{MS}$ masses of the charm and strange quarks, $m_{c}(m_c)=(1.275\pm0.025)\,\rm{GeV}$
 and $m_s(\mu=2\,\rm{GeV})=(0.095\pm0.005)\,\rm{GeV}$,
 from the Particle Data Group \cite{PDG}, just like in our previous works \cite{WangHuangTao-3900,Wang-Hidden-charm}.
In addition,  we take into account  the energy-scale dependence of  all the input parameters at the quark level, such as the quark condensates $\langle\bar{q}q \rangle$, $\langle\bar{s}s \rangle$, the mixed quark condensates $\langle\bar{q}g_s\sigma G q \rangle$, $\langle\bar{s}g_s\sigma G s \rangle$ and the $\overline{MS}$ masses $m_c(\mu)$, $m_s(\mu)$ in the light of   renormalization group equation \cite{Narison-mix},
 \begin{eqnarray}
 \langle\bar{q}q \rangle(\mu)&=&\langle\bar{q}q\rangle({\rm 1 GeV})\left[\frac{\alpha_{s}({\rm 1 GeV})}{\alpha_{s}(\mu)}\right]^{\frac{12}{33-2n_f}}\, , \nonumber\\
 \langle\bar{s}s \rangle(\mu)&=&\langle\bar{s}s \rangle({\rm 1 GeV})\left[\frac{\alpha_{s}({\rm 1 GeV})}{\alpha_{s}(\mu)}\right]^{\frac{12}{33-2n_f}}\, , \nonumber\\
 \langle\bar{q}g_s \sigma Gq \rangle(\mu)&=&\langle\bar{q}g_s \sigma Gq \rangle({\rm 1 GeV})\left[\frac{\alpha_{s}({\rm 1 GeV})}{\alpha_{s}(\mu)}\right]^{\frac{2}{33-2n_f}}\, ,\nonumber\\
  \langle\bar{s}g_s \sigma Gs \rangle(\mu)&=&\langle\bar{s}g_s \sigma Gs \rangle({\rm 1 GeV})\left[\frac{\alpha_{s}({\rm 1 GeV})}{\alpha_{s}(\mu)}\right]^{\frac{2}{33-2n_f}}\, ,\nonumber\\
m_c(\mu)&=&m_c(m_c)\left[\frac{\alpha_{s}(\mu)}{\alpha_{s}(m_c)}\right]^{\frac{12}{33-2n_f}} \, ,\nonumber\\
m_s(\mu)&=&m_s({\rm 2GeV} )\left[\frac{\alpha_{s}(\mu)}{\alpha_{s}({\rm 2GeV})}\right]^{\frac{12}{33-2n_f}}\, ,\nonumber\\
\alpha_s(\mu)&=&\frac{1}{b_0t}\left[1-\frac{b_1}{b_0^2}\frac{\log t}{t} +\frac{b_1^2(\log^2{t}-\log{t}-1)+b_0b_2}{b_0^4t^2}\right]\, ,
\end{eqnarray}
  where $t=\log \frac{\mu^2}{\Lambda^2}$, $b_0=\frac{33-2n_f}{12\pi}$, $b_1=\frac{153-19n_f}{24\pi^2}$, $b_2=\frac{2857-\frac{5033}{9}n_f+\frac{325}{27}n_f^2}{128\pi^3}$,  $\Lambda=213\,\rm{MeV}$, $296\,\rm{MeV}$  and  $339\,\rm{MeV}$ for the quark flavor numbers  $n_f=5$, $4$ and $3$, respectively  \cite{PDG}.
In the present work, we investigate   the hidden-charm tetraquark states $c\bar{c}u\bar{s}$ with strangeness,  it is better to adopt  the quark flavor numbers $n_f=4$, then  evolve all the input parameters at the quark level to a typical energy scale $\mu$, which satisfies the restriction of the energy scale formula  $\mu=\sqrt{M_{Z}-(2{\mathbb{M}}_c)^2}$ with the updated  effective charm quark mass  ${\mathbb{M}}_c=1.82\,\rm{GeV}$ \cite{Wang-Hidden-charm,Wang-tetra-formula,WangZG-eff-Mc,Wang-tetra-IJMPA}. If we take the $Z_{cs}(3985)$ as the ground state tetraquark candidate  for the $Z_{N/P}$ with hidden-charm, with strangeness, and with $J^P=1^+$, the best energy scales (or our preferred energy scales) of the spectral densities at the quark level are $\mu=1.6\,\rm{GeV}$.

Now let us take a short digression to discuss the energy scale dependence of the QCD sum rules. In preforming the operator product expansion, we can choose any energy scale if the perturbative calculations  are  feasible at this special energy scale, and the physical quantities extracted from the QCD sum rules should be  independent  on selections of the energy scales. In this sense, the correlation functions $\Pi(p^2)$ are independent on the energy scales, $\frac{d}{d \mu}\Pi(p^2)=0$.

On the other hand, the two-quark (three-quark, four-quark, $\cdots$) currents $J(x)$ are operators and are renormalized at special energy scales, thus  they are defined at special energy scales and are energy scale dependent quantities,
\begin{eqnarray}
J(x,\mu)&=&L^{\gamma_{J}} J(x,\mu_0)\, ,
\end{eqnarray}
where $L=\frac{\alpha_s(\mu_0)}{\alpha_s(\mu)}$, and the $\gamma_J$ are the anomalous dimensions of the currents $J(x)$.
We usually neglect renormalization of the hadron states, and set the anomalous dimensions of the pole residues (or decay constants) to be  the anomalous dimensions of the current operators,
\begin{eqnarray}
\langle 0|J(0,\mu)|H(p)\rangle&=&\lambda_H(\mu)\, , \nonumber\\
                              &=&L^{\gamma_{J}}\langle 0| J(0,\mu_0)|H(p)\rangle\nonumber\\
                              &=&L^{\gamma_{J}} \lambda_H(\mu_0)\, .
\end{eqnarray}

In fact, even in the QCD sum rules for the conventional $D$ meson, where the radiative corrections of the perturbative terms have been calculated up to the order $\mathcal{O}(\alpha_s^2)$ and the radiative  corrections of the quark condensate have been calculated up to the order $\mathcal{O}(\alpha_s^1)$ \cite{Three-loop-1,Three-loop-2,Two-loop-quark}, the relation for the decay constant $f_D(\mu)=L^{\frac{12}{25}}f_D(\mu_0)$ cannot take account of the energy scale dependence of the QCD sum rules in a consistent way.

On one hand, the non-local operators $J(x)J(0)$ have their own anomalous dimensions $\gamma_{JJ}$, $\gamma_{JJ}\neq 2\gamma_J$, just like re-normalization  of the quark fields $q(x)$ alone is not enough for the conventional current operators $J(x)=\bar{q}(x)\Gamma q(x)$, where the $\Gamma$ stand for some Dirac $\gamma$-matrixes.
On the other hand, we usually neglect the radiative corrections due to the cumbersome calculations and factorize the higher dimensional vacuum condensates to the lower dimension vacuum condensates in performing the operator product expansion, the energy scale dependence of the QCD sum rules is modified. Furthermore, we introduce the continuum threshold parameters  $s_0$ to exclude the contaminations of the higher resonances and continuum states, the  correlation between the thresholds and continuum thresholds is not clear.

The energy scale dependence cannot be absorbed into the pole residues alone, we cannot obtain energy scale independent QCD sum rules for the masses,  selections of the energy scales of the QCD spectral densities influence  the masses extracted from the QCD sum rules.
We can rewrite the energy scale formula in another  form,
\begin{eqnarray}\label{formula-Regge}
M^2_{X/Y/Z}&=&\mu^2+C\, ,
\end{eqnarray}
where the constants $C=4{\mathbb{M}}_c^2$, then  explain  the energy scale formula in another way. We conjecture that the predicted tetraquark  masses and  pertinent energy scales of the QCD spectral densities have a
 Regge-trajectory-like relation, see  Eq.\eqref{formula-Regge}, where the $C$ are free parameters and fitted by the QCD sum rules. Direct calculations have proven that the $C$  have universal values and work well for all the tetraquark (molecular) states.

The standard values of the quark condensates and mixed condensates determined in the original  works still survive \cite{SVZ79,Ioffe-NPB-1981,Ioffe-mixcondensate,ColangeloReview}, while the value of the gluon condensate was updated from time to time in the literatures, however, the standard value determined in the original works \cite{SVZ79} is still feasible \cite{ColangeloReview}. The gluon condensate is  the vacuum expectation value  of the gluon operator  of the order $\mathcal{O}(\alpha_s^1)$, and plays a minor important role in the present work,  the standard value and updated value do not make  much difference. The most important parameter is the $c$-quark mass $m_c(m_c)$.
In 2006, R. D. Matheus et al investigated  the $X(3872) $ as the  diquark-antidiquark type  tetraquark state  with the QCD sum rules  by carrying out the operator product expansion up to the vacuum condensates  of dimension 8 \cite{Narison-3872}.  Thereafter   the QCD sum rules became a powerful theoretical approach in studying the exotic $X$, $Y$ and $Z$ states. In Ref.\cite{Narison-3872}, the $\overline{MS}$ mass  $m_c(m_c) =1.23 \pm 0.05 \rm {GeV}$ was chosen, thereafter the value was adopted without  or with  slightly modified uncertainty  \cite{Nielsen-JPG}. Only in recent years,  new values $m_c(m_c)=1266 \pm 6\,\rm{MeV}$ \cite{Narison-X2900} and $1.275\pm0.025\,\rm{GeV}$ \cite{Zcs-CFQiao} and  $1.275^{+0.025}_{-0.035}\,\rm{GeV}$ \cite{Zcs-molecule-6} were chosen.
The values of the $\overline{MS}$ mass of the $c$-quark listed in { \it  The Review of Particle Physics} in 2012, 2014, 2016, 2018 and 2020   were $1.275\pm0.025\,\rm{GeV}$, $1.275\pm0.025\,\rm{GeV}$, $1.27 \pm 0.03\,\rm{GeV}$, $1.275^{+0.025}_{-0.035}\,\rm{GeV}$ and $1.27 \pm 0.02\,\rm{GeV}$, respectively. We fitted  the constants $C$ with the value  $1.275\pm0.025\,\rm{GeV}$ from  { \it  The Review of Particle Physics (2012)}   \cite{Wang-tetra-formula,WangZG-eff-Mc}, and adopted  the value  eversince.

The energy gaps between the ground states (or 1S) and the first radial excited states (or 2S) of the hidden-charm tetraquark states without strangeness or with hidden-strangeness  are about
  $0.6\,\rm{GeV}$ \cite{Wang-Hidden-charm,Maiani-Z4430-1405,WangZG-X4500,WangZG-axial-Z4600},  we  adopt the continuum threshold parameters as $\sqrt{s_0}=M_{Z}+0.55\pm0.10\,\rm{GeV}=4.55\pm10\,\rm{GeV}$ tentatively, and vary the Borel parameters $T^2$ to satisfy the two
requirements that the pole contributions dominate the QCD sum rules at the hadron  side and the operator product expansions converge rather  quickly at the QCD side
  via trial  and error. In doing so, we should bear in mind that if our numerical results do no support identifying  the $Z_{cs}(3985)$ to be the axialvector tetraquark state, we should refit the best continuum threshold parameters and the best energy scales of the spectral densities at the quark level to obtain the real ground state tetraquark masses.

\begin{figure}
 \centering
 \includegraphics[totalheight=7cm,width=9cm]{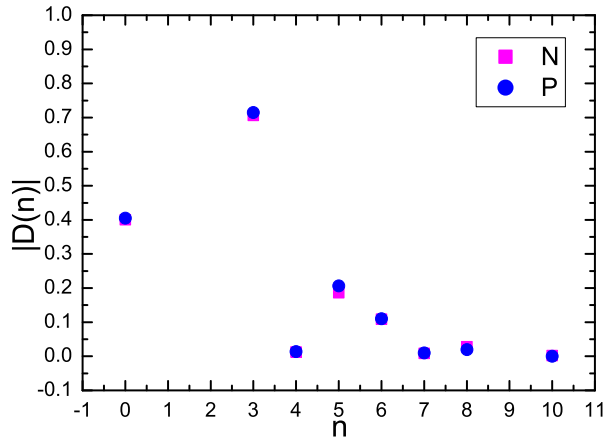}
 \caption{ The absolute values of the contributions of the vacuum condensates in the condition of central values of the input parameters, where the $N$ and $P$ stand for the negative and positive charge conjugation, respectively.  }\label{OPE-n}
\end{figure}

  In the end, we  acquire  the Borel windows (or working Borel parameters) and pole contributions in the two QCD sum rules for the currents $J_\mu^N(x)$ and $J^P_\mu(x)$, and show them plainly  in the Table  \ref{Borel-mass}. From the table, we can see clearly  that the contributions of the lowest pole terms  are about $(42-62)\%$, while the central values exceed $50\%$, so we can say confidently that  the contributions of the  pole terms dominate the QCD sum rules  at the hadron side, one of the fundamental criterions  is  satisfied very well. In Fig.\ref{OPE-n}, we plot the absolute values of the contributions of the vacuum condensates with the centroids  of the  values of all the input parameters under the condition that the total contributions are normalized to be 1.
  From the figure, we can see clearly that the largest contributions  come from the quark condensates $\langle\bar{q}q\rangle$ and $\langle\bar{s}s\rangle$, the vacuum condensates of the dimensions $7$, $8$ and $10$ play a tiny role, the operator product expansion converges very good. Now we can acquire the conclusion confidently that it is reliable and reasonable to extract the tetraquark masses in the Borel windows.

\begin{table}
\begin{center}
\begin{tabular}{|c|c|c|c|c|c|c|c|}\hline\hline
 $J^{PC}$  &$T^2(\rm{GeV}^2)$   &$\sqrt{s_0}(\rm{GeV})$  &$\mu(\rm{GeV})$  &pole          &$M(\rm{GeV})$  &$\lambda(\rm{GeV}^5)$ \\ \hline

$1^{+-}$   &$3.0-3.4$           &$4.55\pm0.10$           &1.6              &$(42-63)\%$   &$3.99\pm0.09$  &$(2.85\pm0.45)\times10^{-2}$   \\ \hline
$1^{++}$   &$3.0-3.4$           &$4.55\pm0.10$           &1.6              &$(41-62)\%$   &$3.99\pm0.09$  &$(2.85\pm0.45)\times10^{-2}$   \\ \hline
\end{tabular}
\end{center}
\caption{ The Borel  windows, continuum threshold parameters, ideal energy scales, pole contributions,   masses and pole residues for the  axialvector
  tetraquark states with  strangeness. } \label{Borel-mass}
\end{table}

\begin{figure}
 \centering
 \includegraphics[totalheight=6cm,width=7cm]{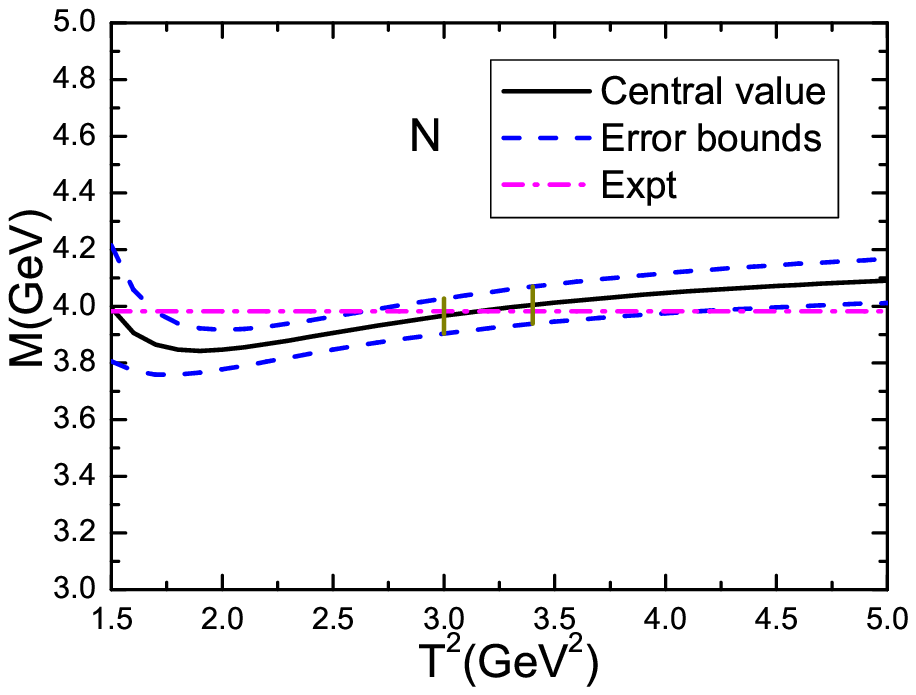}
 \includegraphics[totalheight=6cm,width=7cm]{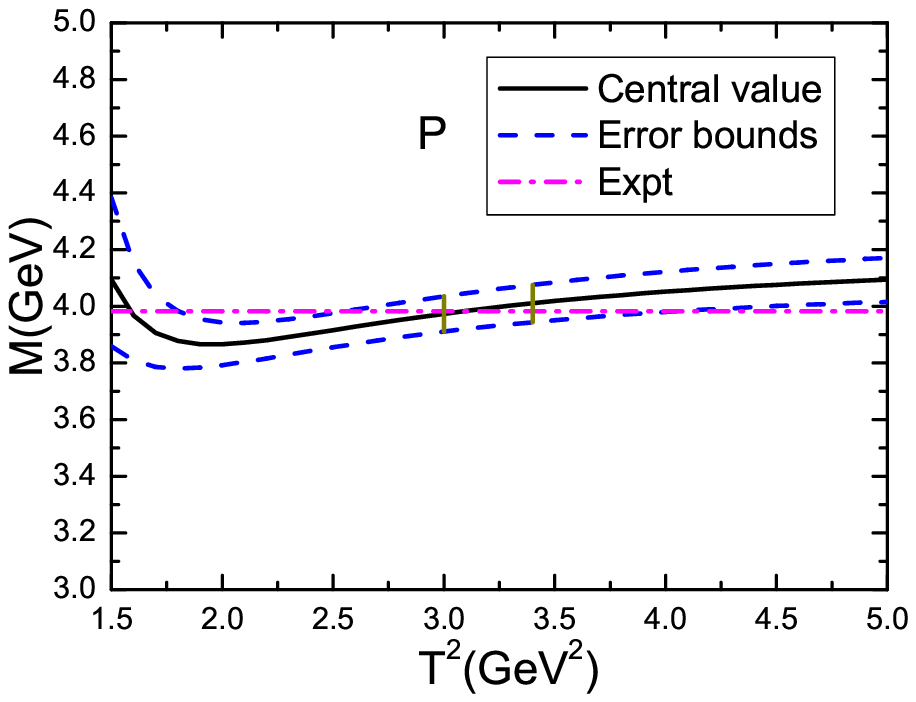}
 \caption{ The masses of the axialvector tetraquark states with variations  of the Borel parameter $T^2$, where the $N$ and $P$ stand for  the negative and positive charge conjugation, respectively, the regions between the two vertical lines are the Borel windows, the Expt stands for  the experimental value of the mass of the $Z_{cs}(3985)$.  }\label{mass-Borel}
\end{figure}

We take into account  all the uncertainties  of the input  parameters  to  accomplish  the error analysis,
and acquire the values of  the masses and pole residues of the  axialvector tetraquark states with hidden-charm and  with the strangeness $S=1$, which are presented  explicitly in Table \ref{Borel-mass} and Fig.\ref{mass-Borel}. From Table \ref{Borel-mass}, we can see clearly that the energy scale $\mu=1.6\,\rm{GeV}$ is consistent with the mass $3.99\,\rm{GeV}$ inferred from  the energy scale formula  $\mu=\sqrt{M_{Z}-(2{\mathbb{M}}_c)^2}$ or the relation between the tetraquark masses and the energy scales of the QCD spectral densities  $M_Z^2=\mu^2+4\mathbb{M}_c^2$. In Fig.\ref{mass-Borel}, we plot the predicted tetraquark masses with respect to variations of the Borel parameters at much larger regions than the Borel windows, furthermore, we also show  the experimental value of the mass of the $Z_{cs}(3985)$ from the BESIII collaboration for the sake of comparing \cite{BES3985}. From the figure, we can see clearly that there really appear very flat platforms in the Borel windows. In the Borel windows, the mass of the $Z_{cs}(3985)$ overlaps with the central values of the masses of the  tetraquark states with strangeness and sharing  the quantum numbers $J^{PC}=1^{+\pm}$. From the Table \ref{Borel-mass}, we can see plainly that the axialvector tetraquark states with the negative and positive charge conjugation have degenerated masses. In fact, the central values of the masses of the axialvector tetraquark states  with the negative and positive charge conjugation are $3.98669\,\rm{GeV}$ and
$3.99303\,\rm{GeV}$ respectively, the axialvector tetraquark states having  the $J^{PC}=1^{++}$ have slightly larger masses than the corresponding states having  the $J^{PC}=1^{+-}$ \cite{Wang-Hidden-charm}. In all the calculations including the present work, we observe that if we choose the same values of the input parameters, such as the quark masses, vacuum condensates, continuum threshold parameters, Borel parameters, etc, the predicted masses of the $J^{PC}=1^{++}$ tetraquark states are slightly larger than that of the $J^{PC}=1^{+-}$ tetraquark states. However, we should bear in mind that the tiny mass difference cannot be quantified considering  the uncertainties of the QCD sum rules.

\begin{figure}
 \centering
 \includegraphics[totalheight=7cm,width=10cm]{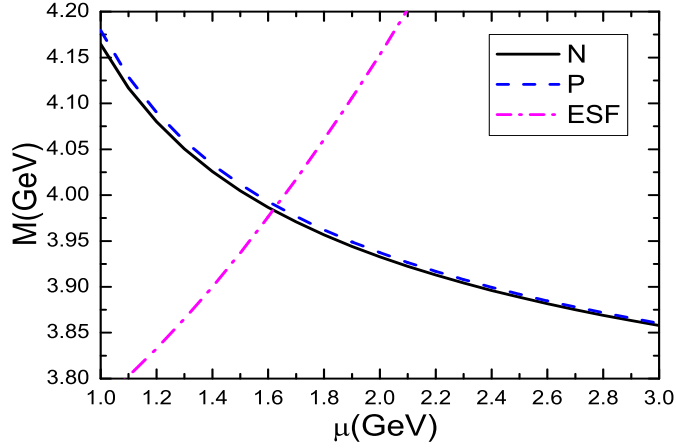}
 \caption{ The masses of the axialvector tetraquark states with variations  of the energy scales $\mu$ in the condition of central values of the input parameters, where the $N$ and $P$ stand for the negative and positive charge conjugation, respectively, the ESF denotes the formula   $M=\sqrt{\mu^2+4\times (1.82\,\rm{GeV})^2}$.  }\label{massmu}
\end{figure}

In Fig.\ref{massmu}, we plot the predicted masses of the $J^{PC}=1^{+-}$ and $1^{++}$ tetraquark states  with variations of the energy scales of the QCD spectral densities in the condition of central values of the input parameters. From the figure, we can see clearly that the predicted masses decrease monotonously with the increase of the energy scales. If we set ${\mathbb{M}}_c=1.82\,\rm{GeV}$ \cite{Wang-Hidden-charm,Wang-tetra-formula,WangZG-eff-Mc,Wang-tetra-IJMPA}, we
can obtain the dash-dotted line $M_{Z}=\sqrt{\mu^2+4\times(1.82\,\rm{GeV})^2}$, which intersects  with the lines of the masses of
the $J^{PC}=1^{+-}$ and $1^{++}$ tetraquark states  at the energy scales about $\mu=1.6\,\rm{GeV}$. In this way, we choose the energy scales of the QCD spectral densities in a consistent way. As a byproduct, we can see clearly, if we choose the same input parameters, the predicted masses of the $J^{PC}=1^{++}$ tetraquark states are slightly larger than that of the $J^{PC}=1^{+-}$ tetraquark states.

When we contract the quark fields in the correlation functions $\Pi_{\mu\nu}(p)$ with Wick theorem, we obtain the formula,
\begin{eqnarray}
\Pi_{\mu\nu}(p)&=&-\frac{i\varepsilon^{ijk}\varepsilon^{imn}\varepsilon^{i^{\prime}j^{\prime}k^{\prime}}\varepsilon^{i^{\prime}m^{\prime}n^{\prime}}}{2}\int d^4x e^{ip \cdot x}   \nonumber\\
&&\left\{{\rm Tr}\left[ \gamma_5C^{kk^{\prime}}(x)\gamma_5 CU^{jj^{\prime}T}(x)C\right] {\rm Tr}\left[ \gamma_\nu C^{n^{\prime}n}(-x)\gamma_\mu C S^{m^{\prime}mT}(-x)C\right] \right. \nonumber\\
&&+{\rm Tr}\left[ \gamma_\mu C^{kk^{\prime}}(x)\gamma_\nu CU^{jj^{\prime}T}(x)C\right] {\rm Tr}\left[ \gamma_5 C^{n^{\prime}n}(-x)\gamma_5 C S^{m^{\prime}mT}(-x)C\right] \nonumber\\
&&\mp{\rm Tr}\left[ \gamma_\mu C^{kk^{\prime}}(x)\gamma_5 CU^{jj^{\prime}T}(x)C\right] {\rm Tr}\left[ \gamma_\nu C^{n^{\prime}n}(-x)\gamma_5 C S^{m^{\prime}mT}(-x)C\right] \nonumber\\
 &&\left.\mp{\rm Tr}\left[ \gamma_5 C^{kk^{\prime}}(x)\gamma_\nu CU^{jj^{\prime}T}(x)C\right] {\rm Tr}\left[ \gamma_5 C^{n^{\prime}n}(-x)\gamma_\mu C S^{m^{\prime}mT}(-x)C\right] \right\} \, ,
\end{eqnarray}
where the $\mp$ correspond  the positive and negative charge conjugations of the currents, respectively,  the $U_{ij}(x)$, $S_{ij}(x)$ and $C_{ij}(x)$ are the full $u$, $s$ and $c$ quark propagators, respectively. In carrying out the operator product expansion, we observe that the dominant contributions come from the first two terms in the bracket, the contributions come from the last two terms in the bracket play a tiny role. It is natural that  we obtain the almost degenerated tetraquark masses.

From the Table \ref{Borel-mass}, we can see clearly that the predicted tetraquark masses $M_Z=3.99\pm0.09\,\rm{GeV}$ are in excellent agreement with the experimental value $3985.2^{+2.1}_{-2.0}\pm1.7\,\rm{MeV}$ from the BESIII collaboration \cite{BES3985}, and support identifying  the  $Z_{cs}(3985)$ to be the   tetraquark state with hidden-charm, with strangeness and  with the $J^{PC}=1^{+-}$. We prefer the quantum numbers  $J^{PC}=1^{+-}$ to the quantum numbers  $J^{PC}=1^{++}$ as the $Z_{cs}(3985)$ was observed in the $D_s^- D^{*0}$ and $D^{*-}_s D^0$ mass spectrum \cite{BES3985}, just like the $Z_c(3900)$, which was observed in the $D\bar{D}^{*}$ mass spectrum \cite{BES-3885}. However, the assignment $J^{PC}=1^{++}$ cannot be excluded as the charge conjugation is not a good quantum number.

In Ref.\cite{Wang-Hidden-charm}, we introduce the four-vector $t^\mu=(1,\vec{0})$ to project the axialvector and vector components of the tensor diquark operators, and   take the color-antitriplet diquark operators $\varepsilon^{ijk}q^{T}_j C\gamma_5 q^{\prime}_k$ ($S$), $\varepsilon^{ijk}q^{T}_j C q^{\prime}_k$ ($P$),
$\varepsilon^{ijk}q^{T}_j C\gamma_\mu q^{\prime}_k$ ($A$),
$\varepsilon^{ijk}q^{T}_j C\sigma^v_{\mu\nu} q^{\prime}_k$ ($\widetilde{A}$),
$\varepsilon^{ijk}q^{T}_j C\gamma_\mu\gamma_5 q^{\prime}_k$ ($V$) and
$\varepsilon^{ijk}q^{T}_j C\sigma^t_{\mu\nu} q^{\prime}_k$ ($\widetilde{V}$)
 as the fundamental   building blocks to construct the four-quark currents to investigate  the mass-spectrum of the   tetraquark states with hidden-charm but without strangeness  in a comprehensive way, where the  $S$, $P$, $A/\widetilde{A}$ and $V/\widetilde{V}$ stand for  the scalar, pseudoscalar, axialvector and vector  diquark operators, respectively,
$\sigma^t_{\mu\nu} =\frac{i}{2}\Big[\gamma^t_\mu, \gamma^t_\nu \Big]$, $\sigma^v_{\mu\nu} =\frac{i}{2}\Big[\gamma^v_\mu, \gamma^t_\nu \Big]$,
$\gamma^v_\mu =  \gamma \cdot t t_\mu$, $\gamma^t_\mu=\gamma_\mu-\gamma \cdot t t_\mu$. In Table \ref{Assignments-Table}, we show plainly  the mass-spectrum of the hidden-charm tetraquark states  $c\bar{c}u\bar{d}$ obtained via the QCD sum rules in Ref.\cite{Wang-Hidden-charm} with the possible assignments.

If we assign the $Z_{cs}(3985)$ to be the cousin of the $Z_c(3900)$
with  strangeness, the mass gap or the light flavor $SU(3)$ mass-breaking effect $M_{Z_{cs}(3985)}-M_{Z_c(3900)}=94\,\rm{MeV}$, then we take the light  flavor $SU(3)$ mass-breaking effect $m_s=0.09\,\rm{GeV}$, and estimate the mass spectrum of the  tetraquark states with hidden-charm and with  strangeness based on our previous work \cite{Wang-Hidden-charm}, which are shown explicitly in Table \ref{Assignments-Zcs-mass}. We should bear  in mind that the charge conjugation is not a very good quantum number in the present case. We cannot exclude that the $Z_{cs}(3985)$ can be identified  as the axialvector hidden-charm tetraquark state with the quantum numbers $J^{PC}=1^{++}$.

\begin{table}
\begin{center}
\begin{tabular}{|c|c|c|c|c|c|c|c|c|}\hline\hline
$Z_c$($X_c$)                                                            & $J^{PC}$  & $M_Z (\rm{GeV})$   & Assignments        &$Z_c^\prime$ ($X_c^\prime$)      \\ \hline

$[uc]_{S}[\overline{dc}]_{S}$                                           & $0^{++}$  & $3.88\pm0.09$      & ?\,$X(3860)$       &       \\

$[uc]_{A}[\overline{dc}]_{A}$                                           & $0^{++}$  & $3.95\pm0.09$      & ?\,$X(3915)$       & \\

$[uc]_{\tilde{A}}[\overline{dc}]_{\tilde{A}}$                           & $0^{++}$  & $3.98\pm0.08$      &                    & \\

$[uc]_{V}[\overline{dc}]_{V}$                                           & $0^{++}$  & $4.65\pm0.09$      &                    & \\

$[uc]_{\tilde{V}}[\overline{dc}]_{\tilde{V}}$                           & $0^{++}$  & $5.35\pm0.09$      &                    &  \\

$[uc]_{P}[\overline{dc}]_{P}$                                           & $0^{++}$  & $5.49\pm0.09$      &                    &  \\ \hline

$[uc]_S[\overline{dc}]_{A}-[uc]_{A}[\overline{dc}]_S$                   & $1^{+-}$  & $3.90\pm0.08$      & ?\,$Z_c(3900)$      &?\,$Z_c(4430)$    \\

$[uc]_{A}[\overline{dc}]_{A}$                                           & $1^{+-}$  & $4.02\pm0.09$      & ?\,$Z_c(4020/4055)$ &?\,$Z_c(4600)$        \\

$[uc]_S[\overline{dc}]_{\widetilde{A}}-[uc]_{\widetilde{A}}[\overline{dc}]_S$     & $1^{+-}$   & $4.01\pm0.09$    & ?\,$Z_c(4020/4055)$ &?\,$Z_c(4600)$     \\

$[uc]_{\widetilde{A}}[\overline{dc}]_{A}-[uc]_{A}[\overline{dc}]_{\widetilde{A}}$ & $1^{+-}$   & $4.02\pm0.09$    & ?\,$Z_c(4020/4055)$ &?\,$Z_c(4600)$    \\

$[uc]_{\widetilde{V}}[\overline{dc}]_{V}+[uc]_{V}[\overline{dc}]_{\widetilde{V}}$ & $1^{+-}$   & $4.66\pm0.10$    & ?\,$Z_c(4600)$      &    \\

$[uc]_{V}[\overline{dc}]_{V}$                                           & $1^{+-}$  & $5.46\pm0.09$      &                    &  \\

$[uc]_P[\overline{dc}]_{V}+[uc]_{V}[\overline{dc}]_P$                   & $1^{+-}$  & $5.45\pm0.09$      &                    &  \\
\hline

$[uc]_S[\overline{dc}]_{A}+[uc]_{A}[\overline{dc}]_S$                   & $1^{++}$  & $3.91\pm0.08$      & ?\,$X(3872)$       &   \\

$[uc]_S[\overline{dc}]_{\widetilde{A}}+[uc]_{\widetilde{A}}[\overline{dc}]_S$     & $1^{++}$   & $4.02\pm0.09$    &?\,$Z_c(4050)$ &   \\

$[uc]_{\widetilde{V}}[\overline{dc}]_{V}-[uc]_{V}[\overline{dc}]_{\widetilde{V}}$ & $1^{++}$   & $4.08\pm0.09$    &?\,$Z_c(4050)$ &    \\

$[uc]_{\widetilde{A}}[\overline{dc}]_{A}+[uc]_{A}[\overline{dc}]_{\widetilde{A}}$ & $1^{++}$   & $5.19\pm0.09$    &               & \\

$[uc]_P[\overline{dc}]_{V}-[uc]_{V}[\overline{dc}]_P$                   & $1^{++}$  & $5.46\pm0.09$      &                    &  \\
\hline

$[uc]_{A}[\overline{dc}]_{A}$                                           & $2^{++}$  & $4.08\pm0.09$      &?\,$Z_c(4050)$      & \\

$[uc]_{V}[\overline{dc}]_{V}$                                           & $2^{++}$  & $5.40\pm0.09$      &                    & \\
\hline\hline
\end{tabular}
\end{center}
\caption{ The possible assignments of the ground state hidden-charm tetraquark states, where the isospin limit is implied \cite{Wang-Hidden-charm}. }\label{Assignments-Table}
\end{table}

\begin{table}
\begin{center}
\begin{tabular}{|c|c|c|c|c|c|c|c|c|}\hline\hline
$Z_c$($X_c$)                                                            & $J^{PC}$  & $M_Z (\rm{GeV})$   & Assignments          \\ \hline

$[uc]_{S}[\overline{sc}]_{S}$                                           & $0^{++}$  & $3.97\pm0.09$      &                           \\

$[uc]_{A}[\overline{sc}]_{A}$                                           & $0^{++}$  & $4.04\pm0.09$      &                     \\

$[uc]_{\tilde{A}}[\overline{sc}]_{\tilde{A}}$                           & $0^{++}$  & $4.07\pm0.08$      &                     \\

$[uc]_{V}[\overline{sc}]_{V}$                                           & $0^{++}$  & $4.74\pm0.09$      &                     \\

$[uc]_{\tilde{V}}[\overline{sc}]_{\tilde{V}}$                           & $0^{++}$  & $5.44\pm0.09$      &                      \\

$[uc]_{P}[\overline{sc}]_{P}$                                           & $0^{++}$  & $5.58\pm0.09$      &                      \\ \hline

$[uc]_S[\overline{sc}]_{A}-[uc]_{A}[\overline{sc}]_S$                   & $1^{+-}$  & $3.99\pm0.09$      & ?\,$Z_{cs}(3985)$        \\

$[uc]_{A}[\overline{sc}]_{A}$                                           & $1^{+-}$  & $4.11\pm0.09$      &         \\

$[uc]_S[\overline{sc}]_{\widetilde{A}}-[uc]_{\widetilde{A}}[\overline{sc}]_S$     & $1^{+-}$   & $4.10\pm0.09$    &     \\

$[uc]_{\widetilde{A}}[\overline{sc}]_{A}-[uc]_{A}[\overline{sc}]_{\widetilde{A}}$ & $1^{+-}$   & $4.11\pm0.09$    &      \\

$[uc]_{\widetilde{V}}[\overline{sc}]_{V}+[uc]_{V}[\overline{sc}]_{\widetilde{V}}$ & $1^{+-}$   & $4.75\pm0.10$    &      \\

$[uc]_{V}[\overline{sc}]_{V}$                                           & $1^{+-}$  & $5.55\pm0.09$      &                      \\

$[uc]_P[\overline{sc}]_{V}+[uc]_{V}[\overline{sc}]_P$                   & $1^{+-}$  & $5.54\pm0.09$      &                     \\
\hline

$[uc]_S[\overline{sc}]_{A}+[uc]_{A}[\overline{sc}]_S$                   & $1^{++}$  & $3.99\pm0.09$      &?\,$Z_{cs}(3985)$         \\

$[uc]_S[\overline{sc}]_{\widetilde{A}}+[uc]_{\widetilde{A}}[\overline{sc}]_S$     & $1^{++}$   & $4.11\pm0.09$    &    \\

$[uc]_{\widetilde{V}}[\overline{sc}]_{V}-[uc]_{V}[\overline{sc}]_{\widetilde{V}}$ & $1^{++}$   & $4.17\pm0.09$    &     \\

$[uc]_{\widetilde{A}}[\overline{sc}]_{A}+[uc]_{A}[\overline{sc}]_{\widetilde{A}}$ & $1^{++}$   & $5.28\pm0.09$    &                \\

$[uc]_P[\overline{sc}]_{V}-[uc]_{V}[\overline{sc}]_P$                   & $1^{++}$  & $5.55\pm0.09$      &                      \\
\hline

$[uc]_{A}[\overline{dc}]_{A}$                                           & $2^{++}$  & $4.17\pm0.09$      & \\

$[uc]_{V}[\overline{dc}]_{V}$                                           & $2^{++}$  & $5.49\pm0.09$      &                  \\
\hline\hline
\end{tabular}
\end{center}
\caption{ The possible assignments of the ground state hidden-charm tetraquark states with strangeness. }\label{Assignments-Zcs-mass}
\end{table}

\section{Conclusion}
In present paper, we choose the scalar and axialvector  diquark operators (in color antitriplet) and antidiquark (in color triplet) operators  as the fundamental    building blocks to construct the four-quark  currents and    investigate  the diquark-antidiquark type  axialvector tetraquark states $c\bar{c}u\bar{s}$ with the QCD sum rules  in the condition that we accomplish the operator product expansion up to the  vacuum condensates of dimension $10$     consistently  based on our reasonable analysis and successful experience, and apply the energy scale formula $\mu=\sqrt{M_{Z}-(2{\mathbb{M}}_c)^2}$ using the effective charmed quark mass ${\mathbb{M}}_c=1.82\,\rm{GeV}$  to fix  the best energy scales of the spectral densities at the quark level. The predicted tetraquark mass $M_Z=3.99\pm0.09\,\rm{GeV}$ is in excellent agreement with the experimental value  $3985.2^{+2.1}_{-2.0}\pm1.7\,\rm{MeV}$ from the BESIII collaboration, and supports identifying  the $Z_{cs}(3985)$   as the cousin of the $Z_c(3900)$ with the quantum numbers $J^{PC}=1^{+-}$. Furthermore, we obtain the mass of the corresponding tetraquark state $c\bar{c}u\bar{s}$ with the quantum numbers $J^{PC}=1^{++}$, which can be compared to the international high energy experimental data in the future.   We take into account the light flavor   $SU(3)$ mass-breaking effect about $90\,\rm{MeV}$, and make crude estimations of  the mass spectrum of the diquark-antidiquark type tetraquark states with hidden-charm and with strangeness.\\
{\bf Note added:} \\
After the  manuscript was submitted to https://arxiv.org/, the LHCb collaboration reported  two new  exotic states with the valence quarks  $c\bar{c}u\bar{s}$  in the $J/\psi  K^+$  mass spectrum  in the decays   $B^+ \to J/\psi \phi K^+$ \cite{LHCb-Zcs4000}.  The most significant state, $Z_{cs}(4000)$, has a mass of $4003 \pm 6 {}^{+4}_{-14}\,\rm{MeV}$, a width of $131 \pm 15 \pm 26\,\rm{MeV}$, and the spin-parity
$J^P =1^+$, while the broader state, $Z_{cs}(4220)$, has a mass of $4216 \pm 24{}^{ +43}_{-30}\,\rm{MeV}$, a width of $233 \pm 52 {}^{+97}_{-73}\,\rm{MeV}$, and the spin-parity $J^P=1^+$ or $1^-$, with a $2\sigma$  difference in favor of the first hypothesis \cite{LHCb-Zcs4000}. As there exist two tetraquark nonets, see Eqs.\eqref{1-nonet}-\eqref{2-nonet}, and there maybe exist mixings between the two tetraquark nonets, so   there are enough rooms to accommodate the $Z_{cs}(3985)$ and $Z_{cs}(4000)$ as the diquark-antidiquark type axialvector tetraquark states with strangeness.

\section*{Acknowledgements}
This  work is supported by National Natural Science Foundation, Grant Number  11775079.


\begin{thebibliography}{99}

\bibitem{BES3985} M. Ablikim et al, Phys. Rev. Lett. {\bf 126} (2021)  102001.

\bibitem{BES-3885}  M. Ablikim et al,   Phys. Rev. Lett. {\bf 112} (2014) 022001.

\bibitem{Ebert:2008kb}  D. Ebert, R. N. Faustov and V. O.Galkin,   Eur. Phys. J.  {\bf C58} (2008) 399.

\bibitem{Ferretti:2020ewe}  J. Ferretti and E. Santopinto,  JHEP  {\bf 04} (2020) 119.




\bibitem{Lee:2008uy}   S. H. Lee, M. Nielsen and U. Wiedner,   J. Korean Phys. Soc.   {\bf 55} (2009) 424.

\bibitem{Dias:2013qga}   J. M. Dias, X. Liu and M. Nielsen,  Phys. Rev.  {\bf D88} (2013)  096014.

\bibitem{SB-Voloshin} M. B. Voloshin, Phys. Lett. {\bf B798} (2019) 135022.


\bibitem{Chen:2013wca}   D. Y. Chen, X. Liu and T. Matsuki,  Phys. Rev. Lett.  {\bf 110} (2013) 232001.

\bibitem{Zcs-CFQiao} B. D. Wan and C. F. Qiao, arXiv:2011.08747 [hep-ph].

\bibitem{Zcs-tetra-quark}  J. Y. Sungu, A. Turkan, H. Sundu and E. V. Veliev, arXiv:2011.13013 [hep-ph].


\bibitem{Zcs-molecule-1} Z. Yang, X. Cao, F. K. Guo, J. Nieves and M. P. Valderrama, arXiv:2011.08725 [hep-ph].

\bibitem{Zcs-molecule-2} M. C. Du, Q. Wang and Q. Zhao, arXiv:2011.09225 [hep-ph].

\bibitem{Zcs-molecule-3} L. Meng, B. Wang and S. L. Zhu, arXiv:2011.08656 [hep-ph].

\bibitem{Zcs-molecule-4} R. Chen and Q. Huang, arXiv:2011.09156 [hep-ph].

\bibitem{Zcs-molecule-5} Z. F. Sun and C. W. Xiao, arXiv:2011.09404 [hep-ph].

\bibitem{Zcs-molecule-6} Q. N. Wang, W. Chen and H. X. Chen, arXiv:2011.10495 [hep-ph].


\bibitem{Zcs-Rescatt-1} J. Z. Wang, D. Y. Chen, X. Liu and T. Matsuki, arXiv:2011.08501 [hep-ph].

\bibitem{Zcs-Rescatt-2} J. Z. Wang, Q. S. Zhou, X. Liu and T. Matsuki, arXiv:2011.08628 [hep-ph].

\bibitem{Zcs-Photopro} X. Cao, J. P. Dai and Z. Yang, arXiv:2011.09244 [hep-ph].



\bibitem{WangHuangTao-3900} Z. G. Wang and T. Huang,  Phys. Rev. {\bf D89} (2014) 054019.

\bibitem{Wang-Hidden-charm} Z. G. Wang,  Phys. Rev. {\bf D102} (2020) 014018.

\bibitem{SVZ79}  M. A. Shifman, A. I. Vainshtein and V. I. Zakharov, Nucl. Phys. {\bf B147} (1979) 385, 448.

\bibitem{PRT85} L. J. Reinders, H. Rubinstein and S. Yazaki, Phys. Rept. {\bf 127} (1985) 1.

\bibitem{WZG-IJMPA-two-par} Z. G. Wang, Int. J. Mod. Phys. {\bf A35} (2020)  2050138.


\bibitem{Wang-tetra-formula}  Z. G. Wang, Eur. Phys. J. {\bf C74} (2014)  2874.

\bibitem{WangZG-eff-Mc} Z. G. Wang, Eur. Phys. J. {\bf C76} (2016) 387.

\bibitem{Wang-tetra-IJMPA} Z. G. Wang and Y. F. Tian, Int. J. Mod. Phys. {\bf A30} (2015) 1550004;
 Z. G. Wang, Commun. Theor. Phys. {\bf 63} (2015) 325;
 Z. G. Wang, Commun. Theor. Phys. {\bf 63} (2015) 466;
 Z. G. Wang, Eur. Phys. J. {\bf C76} (2016)  142.

\bibitem{WangXW-Wang} X. W. Wang and Z. G. Wang, in preparation.


\bibitem{Ioffe-NPB-1981} B. L. Ioffe, Nucl. Phys. {\bf B188} (1981) 317; Erratum: Nucl.Phys. {\bf B191} (1981) 591.

\bibitem{Ioffe-mixcondensate} V. M. Belyaev and B. L. Ioffe, Sov. Phys. JETP {\bf 56} (1982) 493.


\bibitem{ColangeloReview} P. Colangelo and A. Khodjamirian, hep-ph/0010175.

\bibitem{PDG}  P. A. Zyla et al,  Prog. Theor. Exp. Phys. {\bf 2020} (2020) 083C01.


\bibitem{Narison-mix} S. Narison and R. Tarrach, Phys. Lett. {\bf 125 B} (1983) 217.





\bibitem{Three-loop-1} K. G. Chetyrkin and M. Steinhauser, Phys. Lett. {\bf B502} (2001) 104.

\bibitem{Three-loop-2} K. G. Chetyrkin and M. Steinhauser, Eur. Phys. J. {\bf C21} (2001) 319.

\bibitem{Two-loop-quark} M. Jamin and B. O. Lange, Phys. Rev. {\bf D65} (2002) 056005;
Z. G. Wang, Phys. Rev. {\bf C92} (2015)  065205.


\bibitem{Narison-3872} R. D. Matheus, S. Narison, M. Nielsen and J. M. Richard, Phys. Rev. {\bf D75} (2007) 014005.


\bibitem{Nielsen-JPG} R. M. Albuquerque, J. M. Dias, K. P. Khemchandani, A. M. Torres, F. S. Navarra, M. Nielsen  and C. M. Zanetti, J. Phys. {\bf G46} (2019) 093002.



\bibitem{Narison-X2900} R. M. Albuquerque, S. Narison, D. Rabetiarivony and G. Randriamanatrika, Nucl. Phys. {\bf A1007} (2021) 122113.




\bibitem{Maiani-Z4430-1405} L. Maiani, F. Piccinini, A. D. Polosa and V. Riquer, Phys. Rev. {\bf D89} (2014) 114010;
 M. Nielsen and F. S. Navarra,  Mod. Phys. Lett. {\bf  A29} (2014) 1430005;
 Z. G. Wang,  Commun. Theor. Phys. {\bf 63} (2015)  325.

\bibitem{WangZG-X4500} Z. G. Wang, Eur. Phys. J. C77 (2017) 78; Z. G. Wang, Eur. Phys. J. A53 (2017) 19.

\bibitem{WangZG-axial-Z4600} Z. G. Wang, Chin. Phys. {\bf C44} (2020) 063105.

\bibitem{LHCb-Zcs4000} R. Aaij   et al, arXiv:2103.01803 [hep-ex].

\end{thebibliography}
\end{document}